\documentclass[aps,pra,twocolumn,groupedaddress,showpacs,floatfix,nofootinbib]{revtex4}

\usepackage{color,soul}

\usepackage{epsfig}
\usepackage{amssymb}
\usepackage{amsmath}

\usepackage{rotfloat}
\usepackage{theorem}
\usepackage{epic}
\usepackage{graphics}
\usepackage{rotating}
\usepackage{placeins}
\usepackage{lscape}
\usepackage{longtable}
\usepackage{dcolumn}
\usepackage[usenames,dvipsnames]{pstricks}

\usepackage{epsfig}
\usepackage{hyperref}

\begin{document}

\newcolumntype{.}{D{.}{.}{-1}}


\newcommand{\del}{\partial}
\newcommand{\beq}{\begin{equation*}}
\newcommand{\eeq}{\end{equation*}}
\newcommand{\be}{\begin{equation}}
\newcommand{\ee}{\end{equation}}
\newcommand{\beqa}{\begin{eqnarray}}
\newcommand{\eeqa}{\end{eqnarray}}
\newcommand{\bea}{\begin{eqnarray}}
\newcommand{\eea}{\end{eqnarray}}
\newcommand{\req}[1]{Eq.\,(\ref{#1})}

\newcommand{\bra}{\langle}
\newcommand{\ket}{\rangle}
\newcommand{\Tr}{{\rm Tr}\,}
\newcommand{\tr}{{\rm Tr}\,}
\newcommand{\s}{\sigma}
\newcommand{\w}{\omega}
\newcommand{\reci}[1]{\frac{1}{#1}}
\newcommand{\half}{\frac{1}{2}}
\newcommand{\emdash}{\hspace{1pt}---\hspace{1pt}}
\newcommand{\volint}[1]{\int \frac{d^4{#1}}{(2\pi)^4} \;}
\newcommand{\volthree}[1]{\int_0^{#1_F} \frac{d^3{#1}}{(2\pi)^3} \;}
\newcommand{\fixminus}{\raisebox{1.5pt}{\mathunderscore}\hspace{0.5pt}}

\newcommand{\putat}[3]{\begin{picture}(0,0)(0,0)\put(#1,#2){#3}\end{picture}}

\hypersetup{
    colorlinks=true,       
    linkcolor=red,          
    citecolor=green,        
    filecolor=magenta,      
    urlcolor=blue           
}

\urlstyle{same}


\title{Proton form-factor dependence of the
finite-size correction to the Lamb shift in muonic hydrogen}



\author{J.~D.~Carroll} 
\email{jcarroll@physics.adelaide.edu.au}
\author{A.~W.~Thomas}

\affiliation{Centre for the Subatomic Structure of Matter (CSSM),
Department of Physics, University of Adelaide, SA 5005, Australia}
\homepage{http://www.physics.adelaide.edu.au/cssm}

\author{J.~Rafelski}
\affiliation{Departments of Physics, University of
  Arizona, Tucson, Arizona, 85721 USA}

\author{G.~A.~Miller}
\affiliation{University of Washington, Seattle, WA 98195-1560 USA}


\date{\today}

\begin{abstract}
  The measurement of the 2{\it P}$_{3/2}^{F=2}$ to 2{\it
    S}$_{1/2}^{F=1}$ transition in muonic hydrogen by Pohl {\it et
    al.}~\cite{Pohl:2010zz} and subsequent analysis has led to the
  conclusion that the rms radius of the proton differs from the
  accepted (CODATA)~\cite{Mohr:2008fa} value by approximately 4$\%$,
  corresponding to a 4.9$\s$ discrepancy. We investigate the
  finite-size effects\emdash in particular the dependence on the shape
  of the proton electric form-factor\emdash relevant to this
  transition using bound-state QED with nonperturbative, relativistic
  Dirac wave-functions for a wide range of idealised
  charge-distributions and a parameterization of experimental data in
  order to comment on the extent to which the perturbation-theory
  analysis which leads to the above conclusion can be confirmed. We
  find no statistically significant dependence of this correction on
  the shape of the proton form-factor.\par
\end{abstract}


\maketitle



\section{Introduction}\label{sec:intro}

The measurement and subsequent analysis of the 2{\it P}$_{3/2}^{F=2}$
to 2{\it S}$_{1/2}^{F=1}$ transition by Pohl {\it et
  al.}~\cite{Pohl:2010zz} concludes that the proton rms charge-radius
is approximately 4$\%$ smaller than previously accepted (as per the
2006 CODATA\footnote{Recently updated:
  \url{http://physics.nist.gov/cgi-bin/cuu/Value?rp}}
value~\cite{Mohr:2008fa}). If accurate, this would indicate that
either QED is an incomplete description of the contributions to the
transition, something has been missed, or something has been
incorrectly calculated.

Following the work of
Refs.~\cite{Borie:2004fv,Martynenko:2006gz,Martynenko:2004bt}, we
focus on one particular contribution to this transition, namely the
finite-size correction to the 2{\it P}$_{1/2}$--2{\it S}$_{1/2}$ Lamb
shift. We report on the form-factor dependence of this correction as
well as the implications for the analysis leading to the proton rms
charge-radius. Discussions of the form-factor dependence of
finite-size effects have recently been reignited by
de~R\'ujula~\cite{DeRujula:2010zk} and this work should settle claims
made in that reference.

\section{Numerical Method}\label{sec:numerical method}

The numerical method used here has recently been summarised in Carroll
{\it et al.}~\cite{Carroll:2011rv} and will not be repeated here. The
highly abridged description is that we use an effective Dirac equation
for a muon (with a reduced mass appropriate to the $\mu$-p
system). This is expected to provide a precise approximation to the
two-particle Bethe-Saltpeter equation, yielding accurate muon
wave-functions for the various potentials studied.

The eigenvalues for each eigenstate can be calculated by inserting the
various potentials into the effective Dirac equation and integrating
iteratively to produce the converged wave-function. Accuracy is
controlled by comparing the converged eigenvalues with those
calculated using a virial theorem and errors are conservatively found
to be $\pm 0.5~\mu$eV.

\section{Proton Finite-Size Corrections}\label{sec:finite}

As a first non-perturbative approximation, the Lamb shift in hydrogen
is calculated using the point-Coulomb and point vacuum polarization
potentials
\bea
\label{eq:C+VP}
\nonumber 
V(r) &=& V_{\rm C} + V_{\rm VP}(r) \\ 
\nonumber
&=& -\frac{Z\alpha}{r}\\
\nonumber
&&-\frac{Z\alpha}{r}\frac{\alpha}{3\pi}\int_4^{\infty}\frac{e^{-m_{e}qr}}{q^2}\ 
\sqrt{1-\frac{4}{q^2}}\left(1+\frac{2}{q^2}\right)d(q^2), \\
&&
\eea
where here $m_e$ represents the electron mass which arises as this is
a consideration of the production of a virtual electron-positron
pair. The momentum integration variable has been changed to $d(q^2)$,
perhaps disguising the fact that the lower cut-off of $q^2=4$
corresponds to $q=2m_e$\emdash the energy required to produce the
pair.

These potentials can be modified to account for the finite-size of
the proton by convoluting the point potential with the proton
charge-distribution. For example, the modification of the Coulomb
potential gives
the Fourier transform of the Coulomb potential in atoms
\bea 
\widetilde{V}(\vec{q\, })=-Z\alpha 4\pi{G_E(\vec{q\, }^2)\over \vec{q\, }^2},
\eea
%
where we note that as the energy transfer to the proton is essentially
negligible, $\vec{q\, }^2$ and the invariant, $Q^2$, are
functionally identical.
%
The coordinate-space potential can then be written in terms of a
three-dimensional Fourier transform of $G_E(\vec{q\, }^2)$:
\be 
\rho(r)\equiv \int {d^3q\over (2\pi)^3}e^{-i\vec{q}\cdot\vec{r}}G_E(\vec{q\, }^2),
\ee
as
%
\bea \label{eq:C}
V_{C}(r) = -\frac{Z\alpha}{r} \to
-Z\alpha\int\frac{\rho(r')}{|\vec{r}-\vec{r\, }'|}\; d^3r.
\label{vf}\eea

Since the potential of Eq.~(\ref{vf}) involves the proton charge-distribution $\rho(r)$\emdash
itself a function of the rms charge-radius\emdash this leads to a
radius-dependent quantity. The dependence on the choice of
 charge-distribution is investigated here.
We enforce that a charge-distribution must satisfy
\be
\label{eq:normalise}
\int \rho(r)\; d^3r = 1,
\ee
and we can investigate the effect of using different forms for
$\rho(r)$. If we use, say, an exponential form (corresponding to a
dipole form-factor, see Eq.\,(\ref{GDipole}))
\be \label{eq:general}
\rho(r) = A e^{-B r} ,
\ee
then we can define the rms charge-radius via a ratio of the moments of the
charge-distribution, as
\be \label{eq:2ndmoment}
\bra r^2\ket = \frac{\displaystyle\int r^2 \rho(r)\;
d^3r}{\displaystyle\int \rho(r)\; d^3r} ,
\ee
for which the value of $A$ in Eq.~(\ref{eq:general}) is arbitrary,
and which can be rearranged such that we arrive at $B=\sqrt{12/\bra
  r_p^2\ket}$ (and $A=B^3/8\pi$ in order to correctly normalise the
distribution). The normalized exponential charge-distribution is then
given by
\be
\rho_{\rm E}(r) = \frac{\eta^3}{8\pi} e^{-\eta r}; \quad \eta =
\sqrt{12/\bra r_p^2\ket}.
\ee

We can, however, perform the same procedure for alternative
charge-distributions and determine the dependence on this choice. For
a Gaussian charge-distribution\emdash corresponding to a Gaussian
form-factor\emdash the normalised form is given by
\be
\rho_{\rm G}(r) = \left(\frac{\eta'}{\pi}\right)^{3/2}e^{-\eta' r^2};
\quad \eta' = 3/2\bra r_p^2\ket.
\ee
Similarly, for a Yukawa charge-distribution\emdash corresponding to a
monopole form-factor\emdash the normalized form is given by
\be
\rho_{\rm Y}(r) = \left(\frac{\eta''^2}{4\pi}\right)e^{-\eta'' r}/r;
\quad \eta'' = \sqrt{6/\bra r_p^2\ket}.
\ee
 
To ensure that we are considering realistic distributions, we also
include in our analysis a charge distribution extracted from a fit to
experimental data of the Sachs electric form factor of the
proton~\cite{Venkat:2010by}~$G_E(Q^2)$ given in that reference by
\be
G_E(Q^2) = 
\frac{1 + q_6\tau + q_{10}\tau^2 + q_{14}\tau^3}
{1 + q_2\tau + q_4\tau^2 + q_8\tau^3 + q_{12}\tau^4 + q_{16}\tau^5}\ ,
\ee
for which the values of $q_i$ are given in Table~\ref{tab:qi}, and for
which $\tau = Q^2/4M_P^2$. This parameterization is constrained at
${\cal O}(Q^2)$ by
%
\be \label{eq:FFfit}
\lim_{Q^2\rightarrow0}G_E({Q\, }^2) = 1-{Q }^2\frac{\bra r_p^2\ket}{6} + {\cal O}({Q\, }^4)
\ee
to reproduce $\sqrt{\bra r_p^2\ket} = 0.878$~fm. The charge
distribution for this form factor is calculated via a Fourier
transform of $G_E$. We will herein refer to this distribution as
`$G_E$ fitted'.

We note several efforts \cite{Jentschura:2010ty,Bawin:2000px} to
include an additional `Darwin-Foldy' (or similarly named) contribution
to the definition of the charge radius beyond that determined in
Eq.~(\ref{eq:2ndmoment}) and used in Eq.~(\ref{eq:FFfit}). We note
that the Darwin-Foldy term was explicitly calculated by Barker and
Glover~\cite{Barker:1955zz} as part of the Breit potential. Together
with the other terms in the Breit potential this is already included
in the recoil correction to the Lamb shift as calculated by previous
authors~(e.g. \cite{Borie:2004fv}) and as such appears in the
complete analysis determining $r_p$ (Line 17 of Table 1 of
Ref.~\cite{Pohl:2010zz} supplementary).

\begin{figure}[t]
  \includegraphics[angle=90,width=0.45\textwidth]{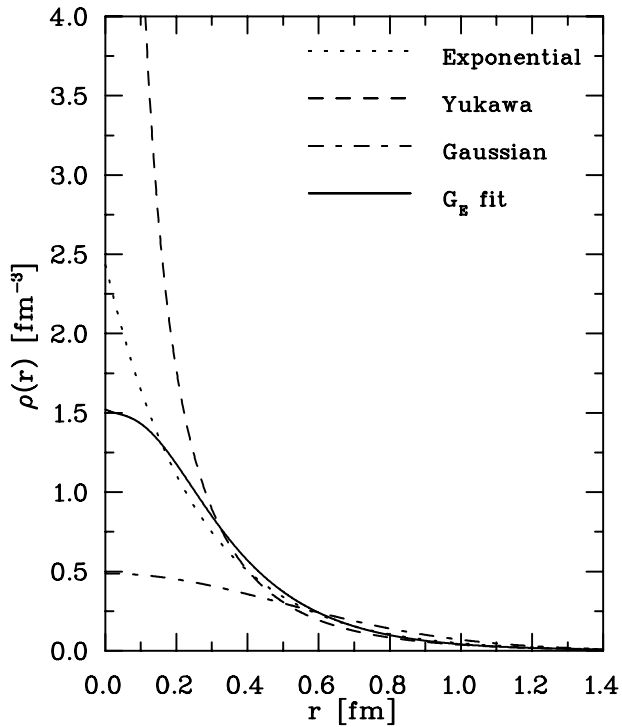}
  \caption{Comparison of exponential, Yukawa, Gaussian, and
    $G_E$-fitted charge-distributions, each normalized to unity as per
    Eq.~(\ref{eq:normalise}), calculated for $\sqrt{\bra r_p^2\ket} =
    0.878~{\rm fm}$. Note the striking differences between the shapes
    below $r=0.8$~fm. The dipole distribution is nearest to the 
     $G_E$-fitted distribution, as expected.\protect\label{fig:rhodata}}
\end{figure}

\begin{figure}[t]
  \includegraphics[angle=90,width=0.45\textwidth]{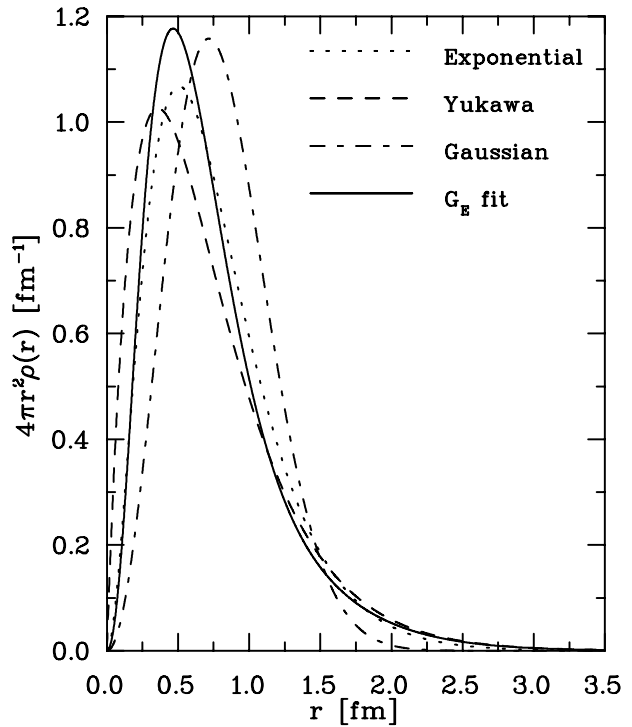}
  \caption{Comparison of exponential, Yukawa, Gaussian, and
    $G_E$-fitted charge-distributions weighted appropriately as they
    contribute to the Lamb shift, each normalized to unity as per
    Eq.~(\ref{eq:normalise}), calculated for $\sqrt{\bra r_p^2\ket} =
    0.878~{\rm fm}$. Here the differences are perhaps not as striking,
    but still noticeably different.\protect\label{fig:4pir2rhodata}}
\end{figure}

\begin{table}[b]
\centering
\caption{\protect\label{tab:qi}Coefficients of polynomial fit to Sachs
  electric form factor data for the proton taken
  from~\cite{Venkat:2010by} as used in
  Eq.~(\ref{eq:FFfit}).\vspace{2mm}}
\begin{ruledtabular}
\begin{tabular}{r.}
\multicolumn{1}{c}{$i$} & \multicolumn{1}{c}{$q_i$} \\[1mm]
&\\[-4mm]
\hline
& \\[-2mm]
 2 & 14.5187212 \\[2mm]
 4 & 40.88333 \\[2mm]
 6 & 2.90966 \\[2mm]
 8 & 99.999998 \\[2mm]
10 & -1.11542229 \\[1mm]
12 & 0.00004579 \\[1mm]
14 & 0.003866171 \\[1mm]
16 & 10.3580447 \\
\end{tabular}
\end{ruledtabular}
\end{table}
The model charge-distributions used in our analysis are plotted for
comparison in Fig.~\ref{fig:rhodata} and we note the striking
differences between the shapes below $r=0.8$~fm. These are plotted
again in Fig.~\ref{fig:4pir2rhodata} where they are weighted by $4\pi
r^2$ 
(as used in the normalization)
and though the differences are reduced, they remain non-trivial.

\begin{figure}[tb]
  \includegraphics[angle=90,width=0.45\textwidth]{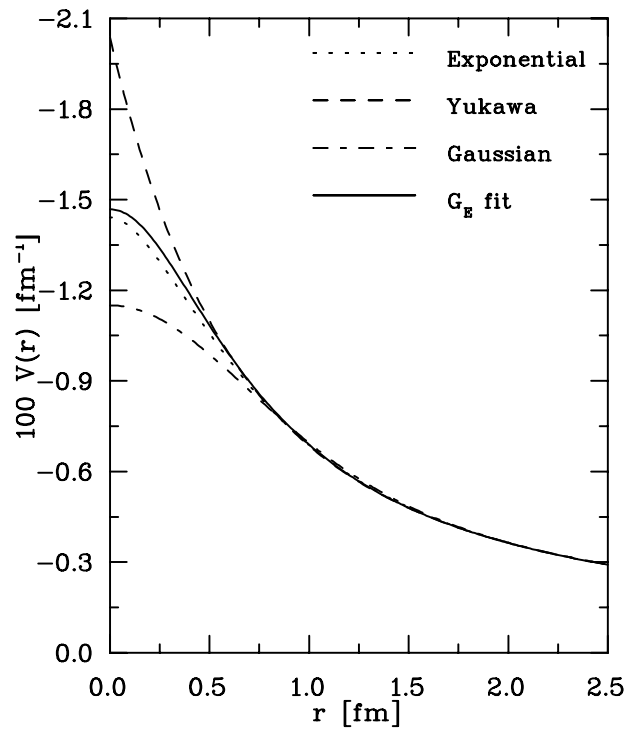}
  \caption{Comparison of exponential, Yukawa, Gaussian, and
    $G_E$-fitted finite-size Coulomb potentials calculated for
    $\sqrt{\bra r_p^2\ket} = 0.878~{\rm fm}$.\protect\label{fig:V}}
\end{figure}

We can use these charge-distributions (calculated at each of four
selected rms charge-radii spanning 0.2~fm surrounding the values given
in Refs.~\cite{Pohl:2010zz} and \cite{Mohr:2008fa}) to calculate the
finite-size Coulomb potential of Eq.~(\ref{eq:C}) \emdash which is
plotted in Fig.~\ref{fig:V}\emdash and hence the converged Dirac
wave-functions in response to this, in order to calculate the
eigen-energies $\lambda_\alpha$ of the 2{\it S}$_{1/2}$ and 2{\it
  P}$_{1/2}$ eigenstates for each charge-distribution. We can then
calculate the deviation of the Lamb shifts ($\delta = \lambda_{2{\it
    S}}-\lambda_{2{\it P}}$) calculated in the point-Coulomb and
finite-Coulomb cases to determine the magnitude of the correction
\be 
\Delta E_{\rm finite} = \delta_{\rm finite} - \delta_{\rm point}.  
\ee

We calculate the proton finite-size correction to the Lamb shift using
the aforementioned effective Dirac equation method for several choices
of charge-distribution (viz exponential, Gaussian, and Yukawa) at
several separated values of the proton rms charge-radius (viz
$\sqrt{\bra r_p^2\ket}$~=~0.7~fm, 0.84184~fm, 0.8768~fm, 0.9~fm). With
this information, we calculate a polynomial fit to the data of
the form
%
%
\be \label{eq:fit}
\Delta E_{\rm finite} = a\bra r^2\ket + b\bra r^2\ket^{3/2},
\ee
in order to compare with other published data.
 A discussion of the role of finite proton size in
vacuum polarization potential is given in
Ref.~\cite{Carroll:2011rv}, Table I.. The relevant parameters of our fits are
shown in Table~\ref{tab:fits}, and we find no significant dependence
on the shape of the proton charge distribution.

Moreover, we are able to make a comparison to the perturbative
finite-size correction (due to the finite-Coulomb potential) to the
Lamb shift as referenced in Ref.~\cite{Pohl:2010zz} and derived in
full in Ref.~\cite{Friar:1978wv} as
\bea
\nonumber
\Delta E_{\rm finite} &=&
-\frac{2\pi}{3}Z\alpha\frac{\left(Z\alpha\mu\right)^3}{2^3\pi}\left[\bra
  r_p^2\ket - \frac{Z\alpha\mu}{2}\bra r_p^3\ket \right.\\
&& \left. +\left(Z\alpha\right)^2\left(F_{\rm Rel} + \mu^2F_{\rm
    NR}\right)\right],
\label{eq:finitecorr}
\eea
with the caveat that this expression was derived for an exponential
charge distribution (corresponding to a dipole form-factor) and does
not equally apply for other distributions.
The derivation of Eq.~(\ref{eq:finitecorr}) assumes that the
Schr\"odinger wave-function is appropriate, in that the value at the
origin
\be
|\phi_n(0)|^2 = \frac{(Z\alpha\mu)^3}{n^3\pi}
\ee
appears, and as such this expression requires a relativistic
correction $F_{\rm Rel}$. Alternatively, by numerically calculating
the converged Dirac wave-functions
we require no such perturbative correction, and a comparison with that
given in Eq.~(\ref{eq:finitecorr}) is consistent. We find agreement
between our Dirac calculation with an exponential distribution and
various evaluations of Eq.~(\ref{eq:finitecorr}) to within $0.05\%$,
as detailed in Table~\ref{tab:fits}.

As the fit to the experimental form factor data is performed at a
single value of $r_p$ we cannot determine a polynomial dependence on
this quantity. We can however interpolate the shifts from our three
models to compare at a single value of $r_p$. The result of such a
comparison is that at $r_p = 0.878$~fm, the contribution to the Lamb
shift due to the finite size of the proton is given by
\bea 
\label{eq:0.878_1}
\Delta E_{\rm finite}^{\rm Exponential}(0.878~{\rm fm}) &=& -3.9850~{\rm
  meV}, \\[3mm]
\label{eq:0.878_2}
\Delta E_{\rm finite}^{\rm Yukawa}(0.878~{\rm fm}) &=& -3.9830~{\rm
  meV}, \\[3mm]
\label{eq:0.878_3}
\Delta E_{\rm finite}^{\rm Gaussian}(0.878~{\rm fm}) &=& -3.9868~{\rm
  meV}, \\[3mm]
\label{eq:0.878_4}
\Delta E_{\rm finite}^{G_E~{\rm fitted}}(0.878~{\rm fm}) &=& -3.9799~{\rm
  meV}, 
\eea
%
in keeping with our conclusion that the form-factor shape is of
negligible influence.

We note Ref.~\cite{Kelly:2004hm} in which a choice of electric
form-factor parameterization is compared to an idealized dipole
form-factor
\be
\label{GDipole}
G_D(Q^2) = \left(1+ Q^2/\Lambda^2\right)^{-2},
\ee
and for which the ratio of the two tends to unity at low $Q^2$. The
ratio remains close to unity up to approximately 1~(GeV/c)$^2$,
re-enforcing that a dipole is a suitable parameterization of the
electric form-factor for the purposes of this analysis.

\begin{table}[b]
\centering
\caption{\protect\label{tab:fits}Coefficients of polynomial fits to
  the finite-size correction to the Lamb shift $\Delta E^{\rm
    Lamb}_{\rm finite}$ (refer to Eq.~(\ref{eq:fit})) in muonic
  hydrogen for several choices of charge-distribution (calculated
  using the finite-Coulomb potential), and selected published
  values. All values in this table include a radiative correction of
  $-0.0275~\bra r_p^2\ket$ as per Ref.~\cite{Pohl:2010zz}. 
  We note that the finite-size effects of the vacuum
  polarization alter these values further, and a discussion of this
  matter can be found in Ref.~\cite{Carroll:2011rv}.\vspace{2mm}}
\begin{ruledtabular}
\begin{tabular}{rc..}
Name & $\rho(r)$ & \multicolumn{1}{c}{$a\ [\propto \bra r^2\ket]$} &  \multicolumn{1}{c}{$b\ [\propto \bra r^2\ket^{3/2}]$} \\[1mm]
&\\[-4mm]
\hline
&\\[-2mm]
exponential & $A e^{-Br}$ & -5.2276 & 0.0351 \\[2mm]
Yukawa & $A e^{-Br}/r$ & -5.2275 & 0.0378 \\[2mm]
Gaussian & $A e^{-Br^2}$ & -5.2276 & 0.0323 \\[2mm]
Ref.~\cite{Pohl:2010zz} & & -5.22495 & 0.0347 \\[1mm]
Ref.~\cite{Borie:2004fv} & & -5.22456 & 0.0346 \\[1mm]
Ref.~\cite{Pachucki:1999zza} & & -5.2249{\phantom{1}} & 0.0363 \\
\end{tabular}
\end{ruledtabular}
\end{table}

\section{Implications for the Proton Radius}

\begin{figure*}[!t]
  \includegraphics[angle=90,width=0.7\textwidth]{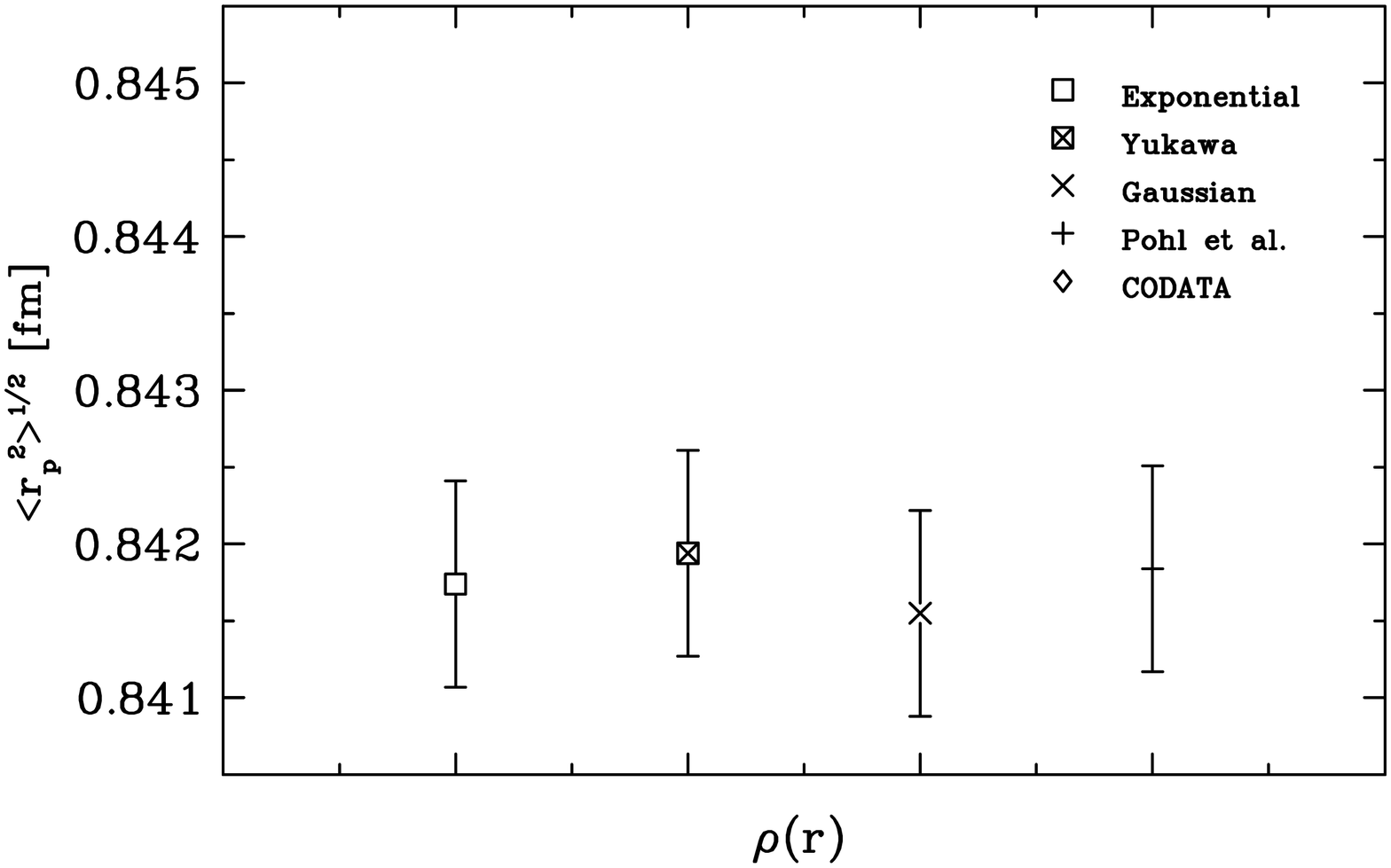}
  \caption{Comparison of calculated proton rms charge-radii (with
    errors) for various charge-distributions. Also shown are the
    values of Pohl et al. from Ref.~\cite{Pohl:2010zz} and (inset) the
    2006 CODATA value, Ref. \cite{Mohr:2008fa}. Error bars are shown
    for all data here, but in many cases they are not visible. It
    should be stressed that the analysis of this data remains reliant
    on the as-yet unknown contribution of off-shell
    effects~\cite{Miller:2011yw}.\protect\label{fig:rdata}}
  \putat{-105}{190}{
    \includegraphics[angle=90,width=0.28\textwidth]{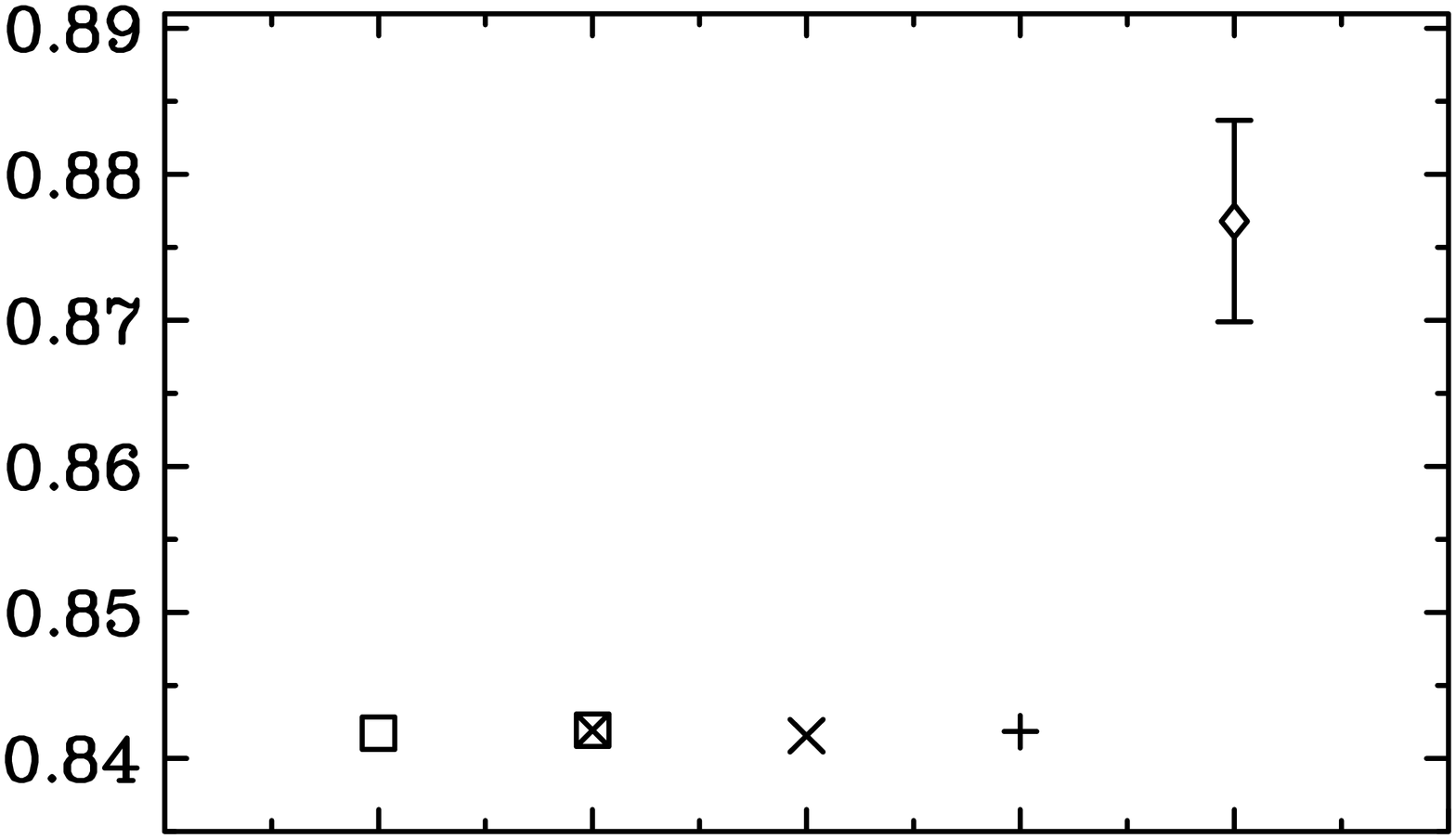}}
\end{figure*}

With the parameterizations of the finite-size contribution to the Lamb
shift determined, it is possible to infer a proton rms charge radius
$\bra r_p^2\ket^{1/2}$ by re-analyzing the measured transition in
muonic hydrogen of Ref.~\cite{Pohl:2010zz}. If we take all other
contributions to the transition {\it prima facie} (we note that
because of the unknown magnitude of off-shell corrections to the
photon-nucleon vertex~\cite{Miller:2011yw} such an analysis is
physically inappropriate) we can solve the cubic equation
\be
\label{eq:cubic}
L_{\rm measured} = L_{\rm r-indep} + a'\bra r^2\ket + b'\bra r^2\ket^{3/2},
\ee 
(where we note that $a'$ and $b'$ may also account for finite-size
effects in the 2{\it S} hyperfine splitting~\cite{Carroll:2011rv}, not
included here, and thus $a'=a$, $b'=b$) in which the measured
transition energy is $L_{\rm measured} = 206.2949\pm 0.0032~{\rm
  meV}$; the sum of theoretical radius-independent contributions to
the transition are $L_{\rm r-indep} = 209.9779\pm 0.0049~{\rm meV}$
(including the full Lamb shift and corrections found
in~\cite{Pohl:2010zz}); and the remaining coefficients are taken from
Table~\ref{tab:fits}. Of the three solutions to Eq.~(\ref{eq:cubic}),
only one is physically meaningful. The physically meaningful value of
the proton rms charge-radius calculated for each of the choices of
charge-distribution are given in Table~\ref{tab:radii} and compared in
Fig.~\ref{fig:rdata}.

We do not calculate a prediction for the proton rms charge radius
based on the $G_E$ fitted charge distribution as the above analysis
requires knowledge of the polynomial dependence on this quantity,
which is absent from this aspect of our investigation. Nonetheless,
the similarity between predictions of the finite-size effect at $r_p =
0.878$~fm as given in Eqs.~(\ref{eq:0.878_1}--\ref{eq:0.878_4})
suggest that no significant changes would be found.

\begin{table}[t]
\centering
\caption{\protect\label{tab:radii}Proton rms charge-radius $\sqrt{\bra
    r_p^2\ket}$ calculated using various charge-distributions. In
  these calculations, the remaining analysis of
  Ref.~\cite{Pohl:2010zz} is taken {\it prima facie}, including the
  radiative (and other) corrections to the finite-size effect. The
  errors in the radii calculated here are dominated by the
  experimental error in $L_{\rm measured}$. Also shown are
  the values obtained in Ref.~\cite{Pohl:2010zz} and the previously
  accepted 2006 CODATA value~\cite{Mohr:2008fa}.\vspace{2mm}}
\begin{ruledtabular}

\begin{tabular}{rc.}
Name & 
$\rho(r)$ & 
\multicolumn{1}{c}{$\sqrt{\bra r_p^2\ket}\ [{\rm fm}]$} \\[1mm]
&\\[-4mm]
\hline
&\\[-2mm]
exponential & $A e^{-Br}$ & 0.84174(67) \\
Yukawa & $A e^{-Br}/r$ &  0.84194(67) \\
Gaussian & $A e^{-Br^2}$ & 0.84155(67)  \\
Ref.~\cite{Pohl:2010zz} & & 0.84184(67) \\
CODATA~\cite{Mohr:2008fa} & & 0.8768(69)\\
\end{tabular}
\end{ruledtabular}
\end{table}

\section{Conclusions}\label{sec:conclusions}

The dependence of the proton rms charge-radius extracted from an
analysis of the measured transition in muonic hydrogen on the choice
of proton charge-distribution (and thus form-factor) investigated here
is shown to be of negligible importance, despite the wide range of
higher-order moments (given that \mbox{$\bra r^4\ket = 5\bra
  r^2\ket^2/2$} for exponential; $5\bra r^2\ket^2/3$ for Gaussian; and
$10\bra r^2\ket^2/3$ for Yukawa charge-distributions) and
investigation of a realistic charge-distribution based on experimental
data (of the electric Sachs form factor).

The analysis of Ref.~\cite{Pohl:2010zz} suggests a discrepancy with
the 2006 CODATA rms charge-radius of 3.99$\%$. Using the
charge-distributions detailed herein, we have calculated a discrepancy
with the 2006 CODATA value of 4.02$\%$ for Gaussian; 4.00$\%$ for
exponential; and 3.98$\%$ for Yukawa charge-distributions, indicating
no significant variation based on this choice alone.

For the purposes of comparison, it is unclear via the references of
Ref.~\cite{Pohl:2010zz} which charge-distribution has been used to
obtain the value found there. We do conclude however that no choice of
charge-distribution is likely to alter the prediction for the proton
rms charge-radius in a statistically significant manner, given the
wide range of shapes investigated here, and the here confirmed 
insensitivity of the muonic hydrogen Lamb shift to the shape
 of the proton charge distribution. 

Turning our attention to the recent estimates of
Ref.~\cite{DeRujula:2010zk} in which the charge-distribution (via the
form-factor) dependence of the results of Ref.~\cite{Pohl:2010zz} are
challenged\emdash we propose that the results found herein quantify
this dependence sufficiently and rule out the shape of the proton
form-factor as a source of the proton radius discrepancy.

%
%

\begin{acknowledgments}
This research was supported in part by the United States Department of
Energy (under which Jefferson Science Associates, LLC, operates
Jefferson Lab) via contract DE-AC05-06OR23177 (JDC, in part); grant
FG02-97ER41014 (GAM); and grant DE-FG02-04ER41318 (JR), and by the
Australian Research Council, FL0992247, and the University of Adelaide
(JDC, AWT). GAM and JR gratefully acknowledge the support and
hospitality of the University of Adelaide while the project was
undertaken.

\par
\end{acknowledgments}
\bibliography{muHrefs}

\begin{thebibliography}{15}
\expandafter\ifx\csname natexlab\endcsname\relax\def\natexlab#1{#1}\fi
\expandafter\ifx\csname bibnamefont\endcsname\relax
  \def\bibnamefont#1{#1}\fi
\expandafter\ifx\csname bibfnamefont\endcsname\relax
  \def\bibfnamefont#1{#1}\fi
\expandafter\ifx\csname citenamefont\endcsname\relax
  \def\citenamefont#1{#1}\fi
\expandafter\ifx\csname url\endcsname\relax
  \def\url#1{\texttt{#1}}\fi
\expandafter\ifx\csname urlprefix\endcsname\relax\def\urlprefix{URL }\fi
\providecommand{\bibinfo}[2]{#2}
\providecommand{\eprint}[2][]{\url{#2}}

\bibitem[{\citenamefont{Pohl et~al.}(2010)\citenamefont{Pohl, Antognini, Nez,
  Amaro, Biraben et~al.}}]{Pohl:2010zz}
\bibinfo{author}{\bibfnamefont{R.}~\bibnamefont{Pohl}},
  \bibinfo{author}{\bibfnamefont{A.}~\bibnamefont{Antognini}},
  \bibinfo{author}{\bibfnamefont{F.}~\bibnamefont{Nez}},
  \bibinfo{author}{\bibfnamefont{F.~D.} \bibnamefont{Amaro}},
  \bibinfo{author}{\bibfnamefont{F.}~\bibnamefont{Biraben}},
  \bibnamefont{et~al.}, \bibinfo{journal}{Nature (and Supplementary Material)}
  \textbf{\bibinfo{volume}{466}}, \bibinfo{pages}{213} (\bibinfo{year}{2010}).

\bibitem[{\citenamefont{Mohr et~al.}(2008)\citenamefont{Mohr, Taylor, and
  Newell}}]{Mohr:2008fa}
\bibinfo{author}{\bibfnamefont{P.~J.} \bibnamefont{Mohr}},
  \bibinfo{author}{\bibfnamefont{B.~N.} \bibnamefont{Taylor}},
  \bibnamefont{and} \bibinfo{author}{\bibfnamefont{D.~B.}
  \bibnamefont{Newell}}, \bibinfo{journal}{Rev. Mod. Phys.}
  \textbf{\bibinfo{volume}{80}}, \bibinfo{pages}{633} (\bibinfo{year}{2008}),
  \eprint{0801.0028}.

\bibitem[{\citenamefont{Borie}(2005)}]{Borie:2004fv}
\bibinfo{author}{\bibfnamefont{E.}~\bibnamefont{Borie}},
  \bibinfo{journal}{Phys. Rev. A} \textbf{\bibinfo{volume}{71}},
  \bibinfo{pages}{032508} (\bibinfo{year}{2005}), \eprint{physics/0410051}.

\bibitem[{\citenamefont{Martynenko}(2008)}]{Martynenko:2006gz}
\bibinfo{author}{\bibfnamefont{A.}~\bibnamefont{Martynenko}},
  \bibinfo{journal}{Phys.Atom.Nucl.} \textbf{\bibinfo{volume}{71}},
  \bibinfo{pages}{125} (\bibinfo{year}{2008}), \eprint{hep-ph/0610226}.

\bibitem[{\citenamefont{Martynenko}(2005)}]{Martynenko:2004bt}
\bibinfo{author}{\bibfnamefont{A.}~\bibnamefont{Martynenko}},
  \bibinfo{journal}{Phys.Rev.} \textbf{\bibinfo{volume}{A71}},
  \bibinfo{pages}{022506} (\bibinfo{year}{2005}), \eprint{hep-ph/0409107}.

\bibitem[{\citenamefont{De~Rujula}(2011)}]{DeRujula:2010zk}
\bibinfo{author}{\bibfnamefont{A.}~\bibnamefont{De~Rujula}},
  \bibinfo{journal}{Phys. Lett.} \textbf{\bibinfo{volume}{B697}},
  \bibinfo{pages}{26} (\bibinfo{year}{2011}), \eprint{1010.3421}.

\bibitem[{\citenamefont{Carroll et~al.}(2011)\citenamefont{Carroll, Thomas,
  Rafelski, and Miller}}]{Carroll:2011rv}
\bibinfo{author}{\bibfnamefont{J.~D.} \bibnamefont{Carroll}},
  \bibinfo{author}{\bibfnamefont{A.~W.} \bibnamefont{Thomas}},
  \bibinfo{author}{\bibfnamefont{J.}~\bibnamefont{Rafelski}}, \bibnamefont{and}
  \bibinfo{author}{\bibfnamefont{G.~A.} \bibnamefont{Miller}},
  \bibinfo{journal}{Phys. Rev.} \textbf{\bibinfo{volume}{A84}}
  (\bibinfo{year}{2011}), \eprint{1104.2971}.

\bibitem[{\citenamefont{Jentschura}(2011)}]{Jentschura:2010ty}
\bibinfo{author}{\bibfnamefont{U.~D.} \bibnamefont{Jentschura}},
  \bibinfo{journal}{Eur. Phys. J.} \textbf{\bibinfo{volume}{D61}},
  \bibinfo{pages}{7} (\bibinfo{year}{2011}), \eprint{1012.4029}.

\bibitem[{\citenamefont{Bawin and Coon}(2001)}]{Bawin:2000px}
\bibinfo{author}{\bibfnamefont{M.}~\bibnamefont{Bawin}} \bibnamefont{and}
  \bibinfo{author}{\bibfnamefont{S.~A.} \bibnamefont{Coon}},
  \bibinfo{journal}{Nucl. Phys.} \textbf{\bibinfo{volume}{A689}},
  \bibinfo{pages}{475} (\bibinfo{year}{2001}), \eprint{nucl-th/0101005}.

\bibitem[{\citenamefont{Barker and Glover}(1955)}]{Barker:1955zz}
\bibinfo{author}{\bibfnamefont{W.~A.} \bibnamefont{Barker}} \bibnamefont{and}
  \bibinfo{author}{\bibfnamefont{F.~N.} \bibnamefont{Glover}},
  \bibinfo{journal}{Phys. Rev.} \textbf{\bibinfo{volume}{99}},
  \bibinfo{pages}{317} (\bibinfo{year}{1955}).

\bibitem[{\citenamefont{Venkat et~al.}(2011)\citenamefont{Venkat, Arrington,
  Miller, and Zhan}}]{Venkat:2010by}
\bibinfo{author}{\bibfnamefont{S.}~\bibnamefont{Venkat}},
  \bibinfo{author}{\bibfnamefont{J.}~\bibnamefont{Arrington}},
  \bibinfo{author}{\bibfnamefont{G.~A.} \bibnamefont{Miller}},
  \bibnamefont{and} \bibinfo{author}{\bibfnamefont{X.}~\bibnamefont{Zhan}},
  \bibinfo{journal}{Phys. Rev.} \textbf{\bibinfo{volume}{C83}},
  \bibinfo{pages}{015203} (\bibinfo{year}{2011}), \eprint{1010.3629}.

\bibitem[{\citenamefont{Kelly}(2004)}]{Kelly:2004hm}
\bibinfo{author}{\bibfnamefont{J.~J.} \bibnamefont{Kelly}},
  \bibinfo{journal}{Phys. Rev.} \textbf{\bibinfo{volume}{C70}},
  \bibinfo{pages}{068202} (\bibinfo{year}{2004}).

\bibitem[{\citenamefont{Pachucki}(1999)}]{Pachucki:1999zza}
\bibinfo{author}{\bibfnamefont{K.}~\bibnamefont{Pachucki}},
  \bibinfo{journal}{Phys. Rev.} \textbf{\bibinfo{volume}{A60}},
  \bibinfo{pages}{3593} (\bibinfo{year}{1999}).

\bibitem[{\citenamefont{Friar}(1979)}]{Friar:1978wv}
\bibinfo{author}{\bibfnamefont{J.~L.} \bibnamefont{Friar}},
  \bibinfo{journal}{Ann. Phys.} \textbf{\bibinfo{volume}{122}},
  \bibinfo{pages}{151} (\bibinfo{year}{1979}).

\bibitem[{\citenamefont{Miller et~al.}(2011)\citenamefont{Miller, Thomas,
  Carroll, and Rafelski}}]{Miller:2011yw}
\bibinfo{author}{\bibfnamefont{G.~A.} \bibnamefont{Miller}},
  \bibinfo{author}{\bibfnamefont{A.~W.} \bibnamefont{Thomas}},
  \bibinfo{author}{\bibfnamefont{J.~D.} \bibnamefont{Carroll}},
  \bibnamefont{and} \bibinfo{author}{\bibfnamefont{J.}~\bibnamefont{Rafelski}},
  \bibinfo{journal}{Phys. Rev.} \textbf{\bibinfo{volume}{A}}
  (\bibinfo{year}{2011}), \eprint{1101.4073}.

\end{thebibliography}
\end{document}